\documentclass[a4paper,12pt]{article}

\usepackage{amsmath}
\usepackage[final]{graphicx}% Include figure files
\usepackage{bm}% bold math
\usepackage{amssymb}
\usepackage{amssymb}
\newcounter{comment}

{\refstepcounter{comment}%
\begin{quote}
\ttfamily\small$\blacksquare$ \textbf{\underline{Comment} $\sharp$\thecomment:}}%
{\end{quote}}

\begin{document}
\hfill
%\begin{minipage}{20ex}\small
%ZAGREB-ZTF-09-02\\
%\end{minipage}

\begin{center}
\baselineskip=2\baselineskip
\textbf{\LARGE{Novel TeV-scale seesaw mechanism with Dirac mediators
}}\\[6ex]
\baselineskip=0.5\baselineskip

{\large Ivica~Picek
%$^{a,}$
\footnote{picek@phy.hr;
% corresponding author
} and
Branimir~Radov\v{c}i\'c
%$^{a,}$
\footnote{bradov@phy.hr}}\\[4ex]
\begin{flushleft}
\it
%$^{a}$
Department of Physics, Faculty of Science, University of Zagreb,
 P.O.B. 331, HR-10002 Zagreb, Croatia\\[3ex]
\end{flushleft}
\today \\[5ex]
\end{center}

\begin{abstract}

We propose novel tree level seesaw mechanism with TeV-scale vectorlike Dirac mediators that
produce Majorana masses of the known neutrinos. The gauge quantum number assignment to the 
Dirac mediators allows them to belong to a weak triplet and a five-plet of nonzero hypercharge.
The latter leads to new seesaw formula $m_{\nu} \sim v^6/M^5$, so that the  empirical 
masses $m_{\nu} \sim 10^{-1}$ eV can be achieved by $M \sim$ TeV new states. There is a limited range of the parameter space with $M \leq$ a few 100 GeV where the tree level contribution dominates 
over the respective loop contributions and the proposed mechanism is testable at the LHC.
We discuss specific signatures for Dirac type heavy leptons produced by Drell-Yan fusion at the LHC.

\end{abstract}
\vspace*{2 ex}

\begin{flushleft}
\end{flushleft}

\clearpage

\section{Introduction}

%\noindent \underline{\it Introduction} :

The existence of neutrino masses represents the first tangible
deviation from the original standard model (SM). In order to account for the smallness of these masses 
one has to modify the particle content of the SM by adding new degrees of freedom. The neutrino masses can be realized through an effective dimension-five operator $LL\Phi\Phi$ \cite{Weinberg:1979sa}  which can be generated both at the tree and the loop level. At the tree level there are only three realizations of the dimension-five operator \cite{Ma:1998dn}: type I \cite{Minkowski:1977sc-etc}, type II \cite{KoK-etc} and type III \cite{Foot:1988aq} seesaw mechanisms, mediated by heavy fermion singlet, scalar triplet and a fermion triplet, respectively. Thereby, the fermionic type I and III models involve Majorana seesaw mediators.
Originally, the seesaw scale was linked to the  high energy GUT scale, while a recent reincarnation of $SU(5)$ GUT model \cite{Bajc:2006ia} provides a low scale hybrid type I and III seesaw model.
In a multiple seesaw approach first put forward in \cite{Ma:2000cc} and later extended in \cite{Grimus:2009mm}, the involved additional fields and additional discrete symmetries 
are instrumental in bringing the seesaw mechanism to the TeV scale \cite{Xing:2009hx} accessible at the Large Hadron Collider (LHC). In this way one avoids  a new hierarchy problem of inexplicable lightness of the Higgs that appears in 
conventional seesaw models.

It is conceivable that fermionic degrees of freedom beyond the three SM generations may be realized as vectorlike Dirac states. The heaviness of the top quark and the lightness of the Higgs boson are the landmarks used in several
extensions of the SM which introduce vectorlike fermions. Recently in \cite{Picek:2008dd} we examined the parameter space
for possible discovery of vectorlike top partner, which by a sort of Dirac seesaw mechanism increases the mass of the top \cite{Vysotsky:2006fx}. Given the importance of understanding the origin of neutrino masses, it makes sense to extend our previous study to leptonic sector and to explore a possible role of TeV-scale vectorlike Dirac fermions as seesaw mediators.

The models with Majorana seesaw mediators, which are bound to have zero
hypercharge, can be extended from the mentioned fermion singlets and triplets
to a five-plet with zero hypercharge.
Such multiplet has been singled out as a viable minimal dark matter (DM)
candidate in \cite{Cirelli:2005uq}. On the other hand, the subject of our study here is a
model with the Dirac fermions as seesaw mediators. This means that we have to invoke
vectorlike fermionic multiplets with nonzero hypercharge. In  contrast to the chiral SM fermions, 
the masses of such vectorlike fermions are not restricted to the electroweak scale.
The large mass of the new fermion multiplets may serve
as a seesaw anchor in a novel seesaw mechanism. In order to have a tree level
seesaw, the newly introduced fermion multiplets have to have a neutral
component which mixes with light neutrinos. New scalar multiplets also
have to have a neutral component which will acquire the vacuum expectation value ({\em vev}). As we will show in
more detail elsewhere \cite{PiR10}, these requirements together with the SM gauge symmetry lead to rather restricted viable quantum number assignments for the new multiplets: the one with vectorlike triplet fermions introduced in \cite{Babu:2009aq}, and the vectorlike fermion five-plet at hand. This vectorlike fermion five-plet along with two additional isospin 3/2 scalar multiplets generates at tree level an effective dimension-nine operator $(LLHH)(H^\dagger H )^2$. Such effective operators have already been studied in \cite{Bonnet:2009ej}. 

At first sight, the proposed seesaw mechanism which leads to Majorana masses for light neutrinos from heavy Dirac fermions, looks very intriguing. We show that by appropriate assignment of lepton numbers, the generation of light neutrino Majorana masses can be traced to lepton number violating  Yukawa interactions. This is in contrast to type I and type III seesaw mechanisms mediated by lepton number violating Majorana masses of heavy fermions.\\

\section{Model with extra fermionic five-plet}

%\noindent \underline{\it Model with fermionic five-plet} :

The model introduced here is  based on the SM gauge group $SU(3)_C
\times SU(2)_L \times U(1)_Y$. In addition to usual SM fermions we introduce a vectorlike Dirac five-plet of leptons, where both left $\Sigma_L = (\Sigma^{+++}_L, \Sigma^{++}_L, \Sigma^+_L, \Sigma^0_L, \Sigma^-_L)$ and right $\Sigma_R = (\Sigma^{+++}_R, \Sigma^{++}_R, \Sigma^+_R, \Sigma^0_R, \Sigma^-_R)$ components transform as
$(1,5,2)$ under the SM gauge group. Also, in addition to  the SM Higgs doublet $H
= (H^+, H^0)$ there are two isospin $3/2$ scalar multiplets $\Phi_1=(\Phi^{0}_1,
\Phi^{-}_1, \Phi^{--}_1, \Phi^{---}_1)$ and $\Phi_2=(\Phi^{+}_2,
\Phi^{0}_2, \Phi^{-}_2, \Phi^{--}_2)$ transforming as $(1,4,-3)$ and $(1,4,-1)$, respectively.

As a prototype model let us consider only one SM lepton doublet $l_L$ which, together with the newly introduced vectorlike lepton five-plet and the two scalar quadruplets, builds the gauge invariant Yukawa terms
\begin{equation}
\mathcal{L}_{\text{Y}} = Y_1 \, \overline l_{L} \Sigma_R
\, \Phi_1  +Y_2 \, \overline \Sigma_L (l_{L})^c \, \Phi^*_2   +\text{H.c.} \ .
\label{Yuk}
\end{equation}
The introduced  vectorlike lepton five-plet has a Dirac mass term
\begin{equation}
\mathcal{L}_\text{mass} =  -\,  M_{\Sigma} \, \overline \Sigma_L \Sigma_R +   \text{H.c.} \ .
\label{Dir}
\end{equation}
An analysis of the scalar potential, carried out later, shows that the neutral components of two scalar quadruplets $\Phi^0_1$ and $\Phi^0_2$ develop the induced {\em vevs} $v_{\Phi_1}$ and $v_{\Phi_2}$, respectively. This means that Yukawa terms from eq.~(\ref{Yuk}) lead to the mass terms connecting the lepton doublet with new vectorlike lepton five-plet.

If we restrict ourself to the neutral components of these multiplets, there are
three neutral left-handed fields $\nu_L$, $\Sigma_L^{0}$ and $(\Sigma_R^{0})^c$ which span
the mass matrix as follows:
\begin{eqnarray}
\mathcal{L}_{\nu \Sigma^0} & = &  \, -\frac{1}{2}
\left(\bar \nu_L \; \overline{\Sigma_L^0} \; \overline{(\Sigma_R^0)^c} \right)
\left( \! \begin{array}{ccc}
0 & m_2 & m_1 \\
m_2 & 0 & M_{\Sigma} \\
m_1 & M_{\Sigma} & 0
\end{array} \! \right) \,
\left( \!\! \begin{array}{c} (\nu_L)^c \\ (\Sigma_L^0)^c \\ \Sigma_R^0 \end{array} \!\! \right)
\; + \mathrm{H.c.}\ .
\label{neutral_mass_matrix}
\end{eqnarray}
Here $m_1$ and $m_2$ result from the first and second term of eq.~(\ref{Yuk}), respectively, and $M_{\Sigma}$ is given by eq.~(\ref{Dir}). Consequently, $m_1$ and $m_2$ are on the scale of the {\em vev} $v_{\Phi_1}$ and {\em vev} $v_{\Phi_2}$ of the neutral components of the scalar quadruplets, but $M_{\Sigma}$ is on the new physics scale $\Lambda_{NP}$ larger then the electroweak scale.
After diagonalizing the mass matrix in eq.~(\ref{neutral_mass_matrix}), the two nearly degenerate heavy Majorana fermions acquire masses which are given by the new scale
$\Lambda_{NP} \sim M_{\Sigma}$, and the light neutrino acquires a Majorana mass
\begin{equation}
m_{\nu} \sim \frac{m_1m_2}{M_\Sigma} \sim \frac {Y_1 Y_2 v_{\Phi_1} v_{\Phi_2}} {M_{\Sigma}} \,,
\label{seesaw}
\end{equation}
that depends on the {\em vevs} of the scalar quadruplets.

Note one important difference in comparison to the usual type I and type III seesaw models where heavy lepton seesaw mediators themselves have Majorana mass terms leading to both light and heavy Majorana fermions. The heavy lepton states in the present model have Dirac mass term, but due to Yukawa terms in eq.~(\ref{Yuk}) the heavy lepton states mix both with $l_L$ and $(l_L)^C$, resulting in Majorana light neutrino. The two heavy Majorana fermions are nearly degenerated (quasi-Dirac), indicating that they stem from one Dirac fermion.

\begin{figure}
\centerline{\includegraphics[scale=1.30]{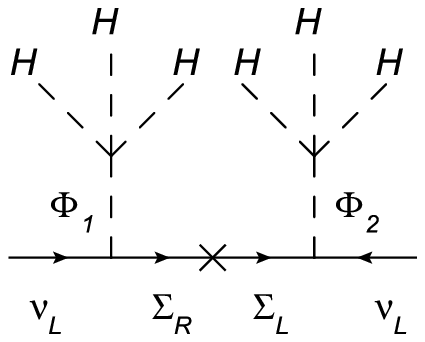}}
\caption{\small Tree level diagram corresponding to dim 9 operator in eq.~(\ref{dim9}). The fermion line flow displays a Dirac nature of the seesaw mediator.}
\label{dim9op}
\end{figure}

The difference between the type I and III seesaw and our seesaw model can also be understood by considering lepton number violation (LNV). Altough there is a freedom in assigning lepton numbers to new multiplets, we can assign for illustrative purposes in all these models (type I, type III and the present one) a non-zero ($L=1$) lepton number to new fermion multiplets and zero lepton number to all scalar multiplets. In our model the lepton number is violated by the second Yukawa term in eq.~(\ref{Yuk}) corresponding to the fermion line flow in Fig. \ref{dim9op}. This is in contrast to type I and III seesaw where the lepton number is violated by Majorana masses of the seesaw mediators. In adopted assignment of lepton numbers the coupling $Y_2$ can have naturally small value in 't Hoft sense \cite{'t Hoft:1979}. There is also attribution of lepton numbers such that LNV in type I and III seesaw comes from the Yukawa terms. However, our vectorlike nonzero hypercharge seesaw messengers do not allow a lepton number assignment that would bring the LNV to the mass term.

Concerning the charged fermions, the triply and doubly charged states do not mix with the SM leptons, while the mixing betwen the standard and new singly charged states is given by
\begin{eqnarray}
\mathcal{L}_{e\Sigma} & = &  \,
\left(\bar e_L \; \overline{\Sigma_L^-} \; \overline{(\Sigma_R^+)^c} \right)
\left( \! \begin{array}{ccc}
m_e & m_3 & m_4 \\
0 & M_{\Sigma} & 0 \\
0 & 0 & M_{\Sigma}
\end{array} \! \right) \,
\left( \!\! \begin{array}{c} e_R \\ \Sigma_R^- \\ (\Sigma_L^+)^c \end{array} \!\! \right)
\; + \mathrm{H.c.} \ .
\label{charged_mass_matrix}
\end{eqnarray}
Here $m_3$ and $m_4$ come from the first and the second term in eq.~(\ref{Yuk}), and $M_{\Sigma}$ is given in eq.~(\ref{Dir}).

Turning to the scalar potential, we restrict ourselves to the following renormalizable terms relevant for our mechanism:
\begin{eqnarray}\label{pot}
V(H, \Phi_1, \Phi_2) &\sim& \mu_H^2 H^\dagger H + \mu^2_{\Phi_1} \Phi^\dagger_1 \Phi_1+ \mu^2_{\Phi_2} \Phi^\dagger_2 \Phi_2 + \lambda_H (H^\dagger H )^2 \nonumber \\
 &+& \{ \lambda_1 \Phi^*_1 H^* H^* H^* + \mathrm{H.c.} \} + \{ \lambda_2 \Phi^*_2 H H^* H^* + \mathrm{H.c.} \} \nonumber\\
 &+& \{ \lambda_3 \Phi^*_1 \Phi_2 H^* H^* + \mathrm{H.c.} \} \ .
\end{eqnarray}
The electroweak symmetry breaking proceeds in the usual way from the {\em vev} $v$ of the Higgs doublet, implying $\mu_H^2<0$. On the other hand, the electroweak $\rho$ parameter dictates the {\em vevs} $v_{\Phi_1}$ and $v_{\Phi_2}$ to be small, implying $\mu^2_{\Phi_1},\mu^2_{\Phi_2}>0$. However, due to the $\lambda_1$ and $\lambda_2$ terms in  eq.~(\ref{pot}) one obtains the induced {\em vevs} for the isospin 3/2 scalar multiplets,
\begin{equation}\label{phivev}
    v_{\Phi_1} \simeq -\lambda_1 \frac{v^3}{\mu^2_{\Phi_1}} \, \, ,\, \, v_{\Phi_2} \simeq -\lambda_2 \frac{v^3}{\mu^2_{\Phi_2}} \, \, .
\end{equation}
These nonvanishing {\em vevs} change the electroweak $\rho$ parameter to $\rho (\Phi_1) \simeq 1-6v^2_{\Phi_1}/v^2$ and $\rho (\Phi_2) \simeq 1+6v^2_{\Phi_2}/v^2$, respectively. If the {\em vevs} are taken separately, the experimental value $\rho=1.0000^{+0.0011}_{-0.0007}$ \cite{PDG08} leads to the upper bounds $v_{\Phi_1}\leq1.9$ GeV and $v_{\Phi_2}\leq2.4$ GeV. Thus, the value of a few GeV can be considered as an upper bound on both $v_{\Phi_1}$ and $v_{\Phi_2}$ if there is no fine-tuning betwen them. By merging eq.~(\ref{seesaw}) and eq.~(\ref{phivev}) we obtain the light neutrino mass
\begin{equation}\label{dim9}
    m_{\nu} \sim \frac {Y_1 Y_2\ \lambda_1\lambda_2\ v^6} {M_{\Sigma}\ \mu^2_{\Phi_1}\ \mu^2_{\Phi_2}} \,,
\end{equation}
which reflects the fact that it is generated from the dimension nine operator shown in Fig. \ref{dim9op}.
\\

\section{Testability}

%\noindent \underline{\it Testability} :

We are able to estimate the high energy scale of our model from eq.~(\ref{dim9}) and the corresponding values for $v_{\Phi_1}$ and $v_{\Phi_2}$ from eq.~(\ref{phivev}). For simplicity let us take the same value for the mass parameters, $\mu_{\Phi_1} \simeq \mu_{\Phi_2} \simeq M_\Sigma \simeq \Lambda_{NP}$. The empirical input values are $v=174$ GeV and $m_\nu\sim0.1$ eV. In one extreme we can assume large values for all couplings $Y_1\sim Y_2\sim \lambda_1\sim \lambda_2\sim1$, resulting in $\Lambda_{NP}\simeq23$ TeV, $v_{\Phi_1}\simeq5$ MeV and $v_{\Phi_2}\simeq4$ MeV. On the other hand, moderate values $Y_1\sim Y_2\sim \lambda_1\sim \lambda_2\sim10^{-2}$ result in $\Lambda_{NP}\simeq580$ GeV and {\em vevs} $v_{\Phi_1}\simeq80$ MeV, $v_{\Phi_2}\simeq60$ MeV still considerably below the few GeV bound.

Effective operators of dimension seven and five generating light neutrino masses are not present at tree level in our model, but they are generated by closing $(H^\dagger H)$ legs in Fig.\ref{dim9op} at one and two loop level, respectively. Because of the characteristic loop suppression factor, these loop generated contributions are smaller than the tree level contribution if $\Lambda_{NP}<4 \pi v\simeq2$ TeV \cite{Gouvea:2008}, which is slightly modified by the number of ways to close the Higgs loop in Fig.\ref{dim9op}. However, there is additional dimension five operator generated at the one loop level by $\lambda_3$ term in eq. (\ref{pot}) giving contribution $Y_1Y_2\lambda_3 v^2/16\pi^2M_\Sigma$. With a reasonable assumption $\lambda_1 \cdot \lambda_2 \simeq \lambda_3$ the ratio of these two contributions is $M^4_\Sigma /16\pi^2  v^4$. This ratio is smaller than one for $\Lambda_{NP} \simeq M_\Sigma  <\sqrt{4 \pi} v \simeq 620$ GeV. Obviously, the later numerical value has to be taken conditionally, and has a meaning of a bound of a few 100 GeV. This bound restricts us to a small part of the parameter space of the model. Since this part is testable at LHC, a non-discovery of the new states under consideration at the LHC would strongly disfavour a tree level generation of neutrino masses in this model. However, such non-discovery would not exclude the loop-level mass generation.

Let us now generalize our initial prototype model by including all three light neutrinos, but still adding only one vectorlike Dirac five-plet. Then the mass matrix in eq.~(\ref{neutral_mass_matrix}) is enlarged to the $5\times5$ matrix. In this case one of the light neutrinos will end up massless at the tree level, but will have a loop generated mass. In the next step, in order to obtain non-zero masses at the tree level for all three light neutrinos, we have to introduce at least two vectorlike Dirac five-plets. Finally, if we want to match the SM pattern of three generations by introducing three vectorlike Dirac five-plets, we will end up with the basis  \{$\nu_{eL},\nu_{\mu L},\nu_{\tau L},\Sigma_{1L},(\Sigma_{1R})^C,\Sigma_{2L},(\Sigma_{2R})^C,\Sigma_{3L},(\Sigma_{3R})^C$\} and the corresponding $9\times9$ mass matrix for the neutral lepton states.

If the tree level contributions to light neutrino masses are dominant, all new states in our model should lie below $\sim$ TeV. Such states are expected to be abundantly produced at colliders through $W$, $Z$ and $\gamma$ Drell-Yan fusion processes. The phenomenology of isospin $3/2$ scalars is beyond the scope of this letter, and provided their masses larger then the mass of fermion five-plet, $\mu_{\Phi_1},\mu_{\Phi_2}>M_\Sigma$, the isospin $3/2$ scalars will not appear in fermion five-plet decays. The associated production of the pairs $(\Sigma^{+++},\overline{\Sigma^{++}})$, $(\Sigma^{++},\overline{\Sigma^{+}})$, $(\Sigma^{+},\overline{\Sigma^{0}})$, $(\Sigma^{0},\overline{\Sigma^{-}})$ via a charged current is a crucial test of the five-plet nature of new leptons. Direct pair production of neutral states $(\Sigma^{0},\overline{\Sigma^{0}})$ via a neutral current, which is not possible for type I and III mediators, is possible for our states with non-zero weak charges.

A recent study \cite{Buckley:2009kv} pointed out that vectorlike fermions of the type considered here are characterized by small mass splitting within a multiplet \cite{Cirelli:2005uq}, and evade observation planed at the LHC with current search strategies. However, this unobservability is avoided for our states which mix with SM particles due to lepton number conserving (LNC) and LNV Yukawa terms in eq.(\ref{Yuk}). Distinctive signatures at colliders could come from a triply charged $\Sigma^{+++}$ decays through an off-shell $\Sigma^{++}$, leading typically to $\Sigma^{+++} \to W^+ W^+ l^+$ decay. On the other hand, doubly charged state decays as $\Sigma^{++} \to W^+ l^+$, and singly charged state decays as $\Sigma^{+} \to W^+ \nu,Z l^+,H^0 l^+$ and $\Sigma^{-} \to W^- \nu,Z l^-,H^0 l^-$. Finally, a neutral seesaw mediator decays as $\Sigma^{0} \to W^\pm l^\mp, Z \nu, H^0 \nu$.

There are two types of LNV processes leading to same sign dileptons and the jets as an appealing signature
\begin{eqnarray}\label{jets}
 q\bar q' \to W^* \to \Sigma^{+} \overline{\Sigma^{0}} &\to& l^+ Z l^+ W^- \to l^+ l^+ jj\ , \nonumber \\
 q\bar q  \to Z^* \to \Sigma^{0} \overline{\Sigma^{0}}  &\to& l^\pm W^\mp l^\pm W^\mp \to l^\pm l^\pm jj\ .
\end{eqnarray}
The process proceeding through the off-shell $W$ boson ends up with one jet having invariant mass of the $W$ boson and the other of the $Z$ boson. Similar processes are also possible in type III seesaw \cite{Bajc:2006ia}. The process proceeding through the off-shell $Z$ boson ends up with both jets having the invariant mass of the $W$ boson. The relative rates of LNV processes with the same sign dileptons with respect to LNC processes with dileptons of opposite sign is determined by the relative strengths of LNC and LNV Yukawa couplings $Y_1$ and $Y_2$.

\section{Conclusions}

%\noindent \underline{\it Conclusions} :

In this letter we propose a novel tree level seesaw mechanism from higher than dim 5 operator mediated by vectorlike five-plet with nonzero hypercharge. In conjunction with the induced {\em vev} of the isospin $3/2$ scalars it generates dim 9 operator for light neutrino masses shown in Fig. \ref{dim9op}. A distinguished feature of the proposed seesaw mechanism is that it leads to Majorana masses for light neutrinos by employing Dirac seesaw mediators. The gauge quantum number assignment for the new states causes that the heavy lepton states mix both with the SM lepton doublet $l_L$ and its charge conjugate $(l_L)^C$, producing light Majorana neutrinos. If the tree level contributions to light neutrino masses are dominant, all new states in our model should lie below $\sim$ TeV and the allowed part of the parameter space of the model will be tested at the LHC.

The model introduced here can be compared to a recent model by Babu et al. \cite{Babu:2009aq}
that uses vectorlike triplet seesaw mediators, which are assumed to be beyond the reach of the LHC. They achieve the neutrino masses suppressed like $m_{\nu} \sim v^4/M^3$ from dim 7 operator, but their main focus is on the phenomenology of triply charged scalar field.  We have complementary focus and new conclusions with respect to them. However, although we have chosen for definiteness the fermion five-plet, our conclusions applay also for nonzero hypercharge triplets in ref. \cite{Babu:2009aq}. 

To our knowledge, we expose for the first time a new type of tree level seesaw mechanism with appealing novel features. We hope that the involved predictive and testable vectorlike fermion five-plet may establish an exciting link between collider phenomenology and the origin of neutrino masses.\\

\section{Acknowledgment}

%\noindent \underline{\it Acknowledgment} :

We thank Goran Senjanovi\'c for critical remarks and B.R. thanks Borut Bajc for useful discussions. Both of us thank Walter Grimus for hospitality offered at the University of Vienna. This work is supported by the Croatian Ministry  of Science, Education and Sports under Contract No. 119-0982930-1016.

\end{document}